\documentclass{elsart}

 \usepackage{epsfig}

\usepackage{amssymb}

\journal{Astroparticle Physics}
\begin{document}

\begin{frontmatter}


\title{Detecting Radio Emission from Cosmic Ray Air Showers and
Neutrinos with a Digital Radio Telescope}
\author{Heino Falcke}


\address{Max-Planck-Institut f\"ur Radioastronomie, Auf dem H\"ugel
69, 53121 Bonn, Germany (hfalcke@mpifr-bonn.mpg.de)}
\author{Peter Gorham}



\address{Dept. of Physics \& Astronomy, Univ. of Hawaii at Manoa, 2505 Correa Rd., Honolulu, HI, 96822, USA (gorham@phys.hawaii.edu)}


\begin{abstract}
We discuss the possibilities of measuring ultra-high energy cosmic rays
and neutrinos with radio techniques.  We review a few of the
properties of radio emission from cosmic ray air showers and show how
these properties can be explained by coherent ``geosynchrotron''
emission from electron-positron pairs in the shower as they move
through the geomagnetic field. This should allow one to use the radio
emission as a useful diagnostic tool for cosmic ray research. A new
generation of digital telescopes will make it possible to study this
radio emission in greater detail.  For example, the planned
Low-Frequency Array (LOFAR), operating at 10-200 MHz, will be an
instrument uniquely suited to study extensive air showers and even
detect neutrino-induced showers on the moon. We discuss sensitivities,
count rates and possible detection algorithms for LOFAR and a
currently funded prototype station LOPES. This should also be
applicable to other future digital radio telescopes such as the
Square-Kilometer-Array (SKA).  LOFAR will be capable of detecting
air-shower radio emission from $>2\cdot10^{14}$ eV to $\sim10^{20}$
eV. The technique could be easily extended to include air shower
arrays consisting of particle detectors (KASCADE, Auger), thus
providing crucial additional information for obtaining energy and
chemical composition of cosmic rays. It also has the potential to
extend the cosmic ray search well beyond an energy of $10^{21}$ eV if
isotropic radio signatures can be found. Other issues that LOFAR can
address are to determine the neutral component of the cosmic ray
spectrum, possibly look for neutron bursts, and do actual cosmic ray
astronomy.
\end{abstract}

\begin{keyword}
Elementary particle processes \sep Radiation mechanisms \sep Radio
telescopes and instrumentation \sep cosmic ray detectors \sep
Interferometry \sep Extensive air showers \sep Cosmic rays
95.30.Cq\sep 
95.30.Gv\sep 
95.55.Jz\sep 
95.55.Vj\sep 
95.75.Kk\sep 
96.40.Pq\sep 
98.70.Sa     
\end{keyword}

\end{frontmatter}

\section{Introduction}

A standard method to observe energetic cosmic rays is simply an array
of particle detectors on the ground measuring either the energetic
muons or electrons in the shower pancake. 
%
%
Since only a small fraction of the total number of particles in the
shower are intercepted by the ground array, the conversion from number
of particles received to primary particle energy is not really
straight forward. Very useful additional information for energy
calibration and particle track recovery of air showers can therefore
be gained by observing radiation emitted from the secondary particles
as the shower evolves. Such radiation is for example Cherenkov
radiation, as observed in the CASA-MIA-DICE experiment, or
fluorescence light from nitrogen molecules in the atmosphere, as seen
by the Fly's Eye detector HiRes and others.  So far this emission is
only detected in the optical and hence requires clear and moonless
dark skies far outside major cities. This gives a duty-cycle of typically 10\%. 

Radio emission might provide an alternative method for doing such
observations including detecting neutrino-induced showers at a higher duty
cycle. This becomes particularly relevant since a new generation of
digital radio telescopes -- designed primarily for astronomical
purposes -- promises a whole new approach to measuring air showers.

\section{Radio properties of extensive air showers}
Radio emission from extensive air showers (EAS) was discovered for the
first time by Jelley and co-workers in 1965 at a frequency of 44
MHz. They used an array of dipole antennas in coincidence with Geiger
counters. The results were soon verified and emission from 2 MHz up to
520 MHz was found in a flurry of activities in the late 1960s. These 
activities ceased almost completely in the subsequent years due to several
reasons: difficulty with radio interference, uncertainty about the
interpretation of experimental results, and the success of other
techniques for air shower measurements.

The radio properties of air showers are summarized in an excellent and
extensive review by Allan (1971). The main result of this review can
be summarized by an approximate formula relating the received
time-integrated voltage of air shower radio pulses to various
parameters, where we also include the presumed frequency scaling:

\begin{eqnarray}\label{crvoltage}
\epsilon_\nu&=&13\, ~\mu {\rm V} {\rm ~m}^{-1} {\rm ~MHz}^{-1}
\left({E_{\rm p}\over 10^{17} ~{\rm eV}}\right)
\left({\sin\alpha\,\cos\,\theta\over\sin 45^\circ\,\cos 30^\circ}\right)\nonumber\\
&\times& \exp \left({-R\over R_0(\nu,\theta)}\right) \left({\nu\over50 ~{\rm
MHz}}\right)^{-1}.
\end{eqnarray}

Here $E_{\rm p}$ is the primary particle energy, $R$ is the offset
from the shower center and $R_0$ is around 110 m, $\theta$ is the
zenith angle, $\alpha$ is the angle of the shower axis with respect to
the geomagnetic field, and $\nu$ is the observing frequency (see also
Allan et al.~1970; Hough \& Prescott 1970). One has to be careful,
however, since in later work by the Haverah Park group consistently
lower values ($1-5 \mu {\rm V} {\rm ~m}^{-1} {\rm ~MHz}^{-1}$ at
$\nu=60$ MHz and $E_{\rm p}=10^{17}$ eV) have been claimed (e.g., Prah
1971; Sun 1975). Some of these discrepancies may be blamed on
sytematic errors in the determination of $E_{\rm p}$ which was used to
normalize the results.

The voltage of the {\it unresolved} pulse in the coherent regime ($\nu
\leq 100~$MHz) can be converted into an ``equivalent flux density'' in
commonly used radio astronomical units, i.e. Jansky\footnote{1 Jy =
10$^{-23}$ erg sec$^{-1}$ cm$^{-2}$ Hz$^{-1}$ = 10$^{-26}$ W m$^{-2}$
Hz$^{-1}$}. Since the conversion of pulsed emission -- which contains
an inherent time scale -- to a flux density is not straightforward, we
define as the equivalent flux density $S_\nu$ of a pulse the flux
density of a {\em steady} continuum source of the same spectrum which
deposits the same energy $E=S_\nu\Delta T\Delta \nu A_{\rm eff}$ in
the antenna during the bandwidth-limited time interval $\Delta t$ as
the pulse. Starting from the Poynting vector, we can define

\begin{equation}\label{fluxconv}
S_{\nu}= \epsilon_\nu^2 \epsilon_0 c/\Delta t= 0.27\,{\rm MJy}\;
\left({\epsilon_\nu\over \mu {\rm V}~ {\rm m}^{-1} {\rm~
MHz}^{-1}}\right)^2 \left({\Delta t\over\mu s}\right)^{-1}.
\end{equation}

The observed pulse duration is $\Delta t\sim1/\Delta \nu$ if the
measurement is bandwidth-limited. In the earlier measurements the
pulses were always unresolved when observing with $\Delta\nu\simeq1$
MHz.

The formula in Eq.~\ref{crvoltage} was determined experimentally from
data in the energy regime $10^{16}\;{\rm eV}\,<E_{\rm p}<10^{18}$
eV. The flux density around 100 MHz seems to depend on primary
particle energy as $S_\nu\propto E_{\rm p}^2$ (Hough \& Prescott 1970;
Vernov et al. 1968; Fig.~\ref{Edependence}) as expected for coherent
emission (see below). This dependency is, however, not yet undoubtedly
established, since a few earlier measurements apparently found
somewhat flatter power-laws (Barker et al. 1967 as quoted in Allan
1971).

\begin{figure}
\centerline{\psfig{figure=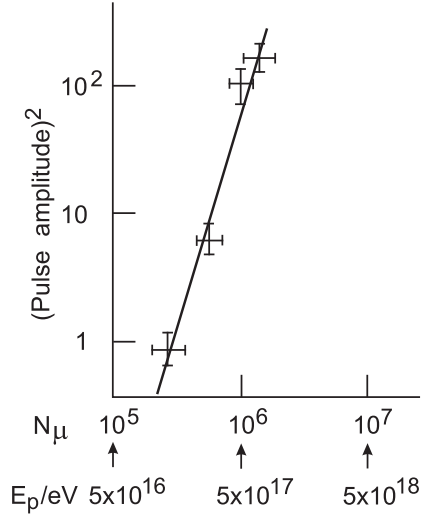,height=0.3\textheight}}
\caption[]{\label{Edependence}The dependence of EAS radio flux on
the primary particle energy as measured by Vernov et al. (1968)
following roughly a $E_{\rm p}^2$ power law. Some earlier papers found
somewhat flatter dependencies.}
\end{figure}

\begin{figure}
\centerline{\psfig{figure=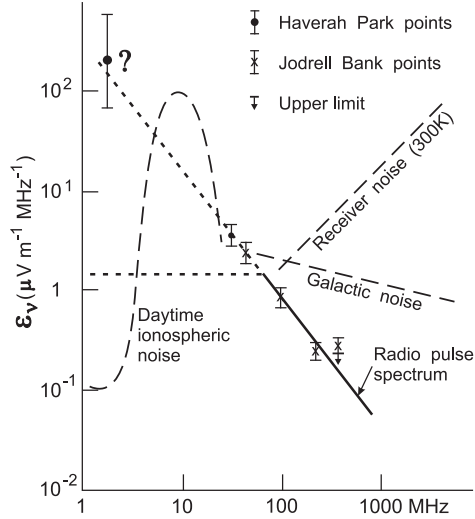,height=0.3\textheight}}
\caption[]{\label{crspectrum}A tentative radio pulse spectrum (from Allan 1971 and Spencer
1969) from 2 MHz to 520 MHz for particles normalized to $E_{\rm
p}=10^{17}$ eV. This has to be squared to get a flux density
spectrum. The data are not simultaneous.  The 2 MHz point was later
questioned and evidence for a flattening of the spectrum (lower
short-dashed line) below 100 MHz was found (e.g. Sun 1975). Various
noise contributions as a function of frequency are also shown.}
\end{figure}

The spectral form of the radio emission was claimed to be valid in the
range 2 MHz $\le\nu\le$ 520 MHz but in general is also fairly
uncertain.  In fact, only very few data on the spectral dependence of
EAS radio emission exist (e.g., Spencer 1969). Figure \ref{crspectrum}
shows a tentative EAS radio spectrum with a $\nu^{-2}$ dependence for
the flux density ($\nu^{-1}$ dependence for the voltage). The 2 MHz
data point was made with a different experiment and there is a real
possibility that the spectrum is actually flat between 10-100 MHz (see
Sun 1975; Datta et al. 2000). The polarization of the emission could
be fairly high and is basically along the geomagnetic E-W direction
(Allan, Neat, \& Jones 1967; Sun 1975) which strongly supports an
emission mechanism related to the geomagnetic field. Most recent
attempts to measure the emission with a single antenna (Green et
al.2002) produced only upper limits consistent with the older
measurements.

\begin{figure}
\centerline{\psfig{figure=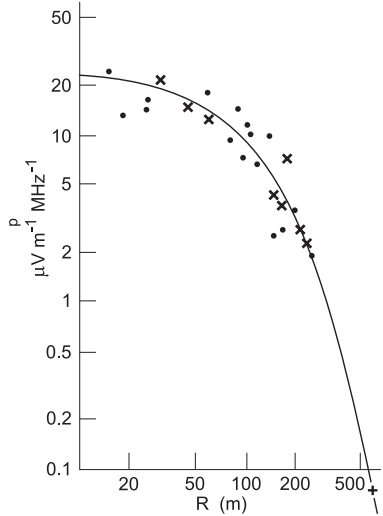,height=0.3\textheight}}
\caption[]{\label{beamshape} Normalized radio pulse amplitudes in
$\mu$V m$^{-1}$ MHz$^{-1}$ at 55 MHz as a function of lateral distance $R$ in
meters from the shower axis. Each data point corresponds to one
measured cosmic ray event. The amplitudes were normalized to a
reference energy of $E_{\rm p}=10^{17}$ eV assuming the above
mentioned linear dependence of voltage on primary particle energy. The
measurements were made for zenith angles $\theta<30^\circ$. Crosses
and dots represent different particle energy bins between $10^{17}$ eV
and $10^{18}$ eV. The plus sign at 500 meters marks a single $10^{19}$
eV event (from Allan 1971).}
\end{figure}

Finally, one needs to consider the spatial structure of the radio
pulse. The current data strongly support the idea that the emission
is not isotropic but is highly beamed in the shower direction. Figure
\ref{beamshape} shows EAS radio pulse amplitude measurements as a
function of distance $R$ from the shower axis -- the flux density
drops quickly with offset from the center of the shower. The
characteristic radius of the beam is of order 100 meter for a
$10^{17}$ eV vertical shower, with the emission presumably originating
at 5-7 km distance above an observer at sea level. The implied angular
diameter of the beam is thus $\Theta \simeq 0.2/6 = 1.9^{\circ}$.

\section{Emission process}

Experiments have clearly established that cosmic ray air showers
produce radio pulses. The original motivation was due to a suggestion
from Askaryan (1962) who argued that annihilation of positrons would
lead to a negative charge excess in the shower, thus producing
Cherenkov radiation as it rushes through the atmosphere. At radio
frequencies the wavelength of the emission is larger than the size of
the emitting region and the emission should be coherent. The radio
flux would then grow quadratically with the number of particles rather than
linearly and thus would be greatly enhanced.  This effect is important in
dense media where it was already experimentally verified (Saltzberg et
al. 2001; see below) and is important for detecting radio emission
from neutrino showers in ice or on the moon.

However, the dependence of the emission on the geomagnetic field
detected in several later experiments indicates that another process
may be important.  The basic view in the late 1960s was that the
continuously created electron-positron pairs were then separated by
the Lorentz force in the geomagnetic field which led to a transverse
current in the shower. If one considers a frame moving along with the
shower, one would observe electrons and positrons drifting in opposite
directions impelled by the transverse electric field induced by the
changing geomagnetic flux swept out by the shower front.  (Only in the
case of shower velocity aligned with the magnetic field lines will
this induced electric field vanish).  This transverse current then
produces dipole (or Larmor) radiation in the frame of the shower. When
such radiation is Lorentz-transformed to the lab frame, the boost then
produces strongly forward-beamed radiation, compressed in time into an
electro-magnetic pulse (EMP).  This was calculated by Kahn \& Lerche
(1966) and also Colgate (1967). Some more involved Monte Carlo
calculations of this process for air showers have been announced by
Dova et al. (1999). In addition there are some claims that the radio
emission could also be influenced by the geoelectric field during
certain times. This was inferred from increased radio amplitudes
associated with EAS during thunder storms (Mandolesi et al. 1973), but
in most regions this should be relatively rare events.

Overall the theoretical basis is still not very well developed and we
feel that for a physical understanding it may be much easier to think
of the emission simply as being coherent synchrotron in the earth's
magnetic field (or yet shorter ``coherent geosynchrotron emission'')
as we show in the following. Coherence is achieved since the shower in
its densest regions is not wider than a wavelength around 100 MHz and
at a few kilometer height the phase shift due to the lateral extent of
the shower for an observer on the ground is similarly less than a
wavelength even out to some 100 meter from the core. Most of the
electrons in the shower are actually concentrated within a region
smaller than this (see Antoni et al. 2001 for measurements of the
lateral distribution on the ground) and here we simply ignore emission
from larger radii.

The proposed geosynchrotron process is probably equivalent to the
previous suggestions since it is derived from the basic formula for
dipole radiation and the Poynting vector but does not require a
consideration of charge separation. The different sign of the charges
of electrons and positrons in the shower is almost completely canceled
by the opposite signs of the Lorentz force acting on electrons and
pairs. Hence both populations will contribute in roughly the same way
to the total flux and will not interfere destructively. To an observer
at the ground the acceleration vectors of electrons and positrons
projected on the sky point in opposite directions and hence the
systems resembles a radiating dipole, with electrons going in one
direction and 'holes' going in the other direction.

The radiated power for a relativistic particle of charge $q$ at the
location $\vec{r}$ can be determined by performing a relativistic
transformation of the Larmor formula from an instantaneous rest frame of
the particle (see for example Rybicki
\& Lightman 1979) to the observer frame. The radiated
power of a particle is given by the second derivative of the dipole
moment and hence the particle's acceleration $\ddot{\vec{r}}$:

\begin{equation}
P_{\rm q}={2 q^2 \ddot{\vec{r}} \cdot \ddot{\vec{r}} \over 3c^3}.
\end{equation}

The acceleration of the charge $q$ with mass $m_{\rm q}$ and Lorentz factor
$\gamma_{\rm q}$ in a magnetic field $\vec{B}$ is given by the Lorentz force

\begin{equation}
\ddot{\vec{r}}={q\over \gamma_{\rm q} m_{\rm q} c}\vec{v}\times\vec{B}.
\end{equation}

Transforming to the observer frame we have $\ddot{\vec{r}}=\gamma_{\rm
q}^2\ddot{\vec{r}}'$ and the above equations can be combined to yield
the {\it emitted} power for synchrotron radiation

\begin{equation}
P_{\rm q}={2 q^2 \over 3c^3} \gamma_{\rm q}^4 {q^2v_\perp^2B^2\over
\gamma_{\rm q}^2 m_{\rm q}^2 c^2}={2 q^4 \over 3c^5m_{\rm q}^2} \gamma_{\rm q}^2 v_\perp^2B^2
\label{power1}
\end{equation}
where $v_\perp$ is the velocity component perpendicular to the
magnetic field and $B=\left|\vec{B}\right|$ . 
In the coherent regime of a shower we could consider $N$
particles of charge $e$ and mass $m_{\rm e}$ acting as a single charged
particle of charge $q=Ne$ and mass $m_{\rm q}=Nm_{\rm e}$, yielding a $N^2$
enhancement over the single-electron power: 

\begin{equation}
P_{\rm N\cdot e}=N^2P_{\rm e}.\label{power2}
\end{equation}

An air shower develops in three stages: the initial rapid buildup via
a multiplicative cascade process, culminating in a broad maximum where
ionization energy losses of the dominant electrons \& positrons
roughly equal their gamma-ray production through bremsstrahlung (at a
critical energy of about 80 MeV in air), then followed by a gradual
decay as the electrons lose energy through ionization.  Early in the
shower development the particle pancake is more compact and
coherence is more complete, while after shower maximum dissipation and
electron straggling reduce the coherence.  Thus most of the radio
flux is produced prior to and within the shower maximum region. This
maximum occurs at column depths of about 550-650 g cm$^{-2}$ for
showers of $10^{17}$ eV, increasing to about 800 g cm$^{-2}$ at
$10^{20}$ eV. As noted above, these depths correspond to heights above
sea level of 5-7 km for $10^{17}$ eV showers from the zenith, but
vertical showers at $10^{20}$ eV are reaching their maximum near sea
level, and the emission thus tends to be produced in the near field
for vertical showers at higher energies.

The broad peak in the electron energy distribution of a typical shower
is at or even below 30 MeV near the shower maximum (Allan
1971)\footnote{Note that the maximum in the electron distribution is
lower than the {\it average} electron energy or the often quoted
characteristic energy.}, i.e. the electron distribution starts to
cut-off below Lorentz factors of $\gamma_{\rm e, min}\sim60$ which we
take as a reference value. At this energy the electrons and positrons
will gyrate around the magnetic field with a gyro radius
\begin{equation}
r_{\rm gyr}={\gamma_{\rm e} m_{\rm e} c^2\over e B}\simeq3.4\, {\rm km}
\left({\gamma_{\rm e}\over60}\right).
\end{equation}
Since the radiation length of electrons in air is about 40 g
cm$^{-2}$, which corresponds to a mean free path for electrons of only
$\sim$ 1 km at 6 km height, the electrons will never complete a full
gyration.

Because of relativistic beaming the radiation will only be visible as
long as the observer is within the beaming cone of full-width opening
angle $\phi=2/\gamma_{\rm e}$. This corresponds to a gyration section
of $s\simeq r_{\rm gyr}\cdot(2/\gamma_{\rm e})\sim 0.1$ km which
is less than the mean free path. The pulse is visible only for a
time interval $\Delta t=t_2-t_1=s/c\,(1-\beta)$. As commonly known in
synchrotron theory (e.g., Rybicki \& Lightman 1979) the $(1-\beta)$
factor accounts for the fact that the relativistically moving emitting
electrons at time $t_2$ will almost have caught up with the photons
emitted at $t_1$, leading to a shortened $\Delta t$ for an observer at
rest. For $\beta\rightarrow1$ this factor expands to
$(1-\beta)=1-\sqrt{1-\gamma^{-2}}\simeq1/2\gamma^2$, yielding

\begin{equation}
\Delta t_{\rm sync}\simeq { m_{\rm e} c\over e \gamma_{\rm e}^2 B}\simeq0.05\,{\rm
ns}\;\left(\gamma_{\rm e}/60\right)^{-2}.
\label{dt1}
\end{equation}

For a single electron the emission would appear as a short pulse with
that duration. The emitted spectrum is the Fourier transform of the
pulse, which is essentially flat up to a maximum frequency of order
$\nu_{\rm sync}\simeq1/\Delta t_{\rm sync}\simeq 19\,{\rm
GHz}\,\left({\gamma_{\rm e}/60}\right)$. For actual synchrotron
radiation the pulse is distinctly non-Gaussian. At low frequencies the
frequency spectrum therefore rises slowly as $\nu^{1/3}$, up to a
characteristic frequency

\begin{equation}\label{nusyncmax}
\nu_{\rm c}={3e\gamma_{\rm e}^2B\over4\pi m_{\rm e} c}=4.5\,{\rm GHz}\,\left({\gamma_{\rm
e}/60}\right)^2.
\end{equation}
The maximum of the frequency spectrum is found at a frequency of
$\sim0.29\nu_{\rm c}$ (see, e.g., Rybicki \& Lightman 1979, Chap. 6).

In an actual shower the pulse duration is further broadened (maximal
bandwidth is limited) by the finite thickness of the emitting layer and
the light travel time. For a typical shower thickness of $z_{\rm
sh}\sim 2$m (e.g. Linsley 1986) in the inner regions around the core
we find

\begin{equation}
\Delta t=z_{\rm sh}/c\simeq 7\,{\rm ns}\left({\Delta z\over 2\,{\rm m}}\right)\Rightarrow\nu_{\rm max}\simeq
150 \,{\rm MHz} \left({\Delta z\over 2\,{\rm m}}\right)^{-1}.
\end{equation}

The shower thickness will widen towards the outer regions and
realistically one could have contributions at different frequencies
from different locations.  The dominant contribution, however, would
still come from the region close to the core and hence our estimate
should be roughly correct. The flatness of the spectrum in the 50 MHz
regime predicted by this simple picture would be consistent with the
later Haverah Park measurements (e.g., Prah 1971) but could not
account for the claimed 2 MHz detection and the backward extrapolation
made by Spencer (1969).

We can now estimate the equivalent flux density (see
Eq.~\ref{fluxconv}) of the geosynchrotron pulse. First we have to
convert the {\it emitted} power (Eqs.~\ref{power1} \& \ref{power2})
into the {\it received} power. For a single pulse we have to take into
account that the time of emission is shortened by a factor
$(1-\beta\cos\theta)\simeq\gamma^{-2}$ for a line-of-sight angle
$\theta=\gamma^{-1}$ and $\beta\rightarrow1$ (see above and the
discussion in Rybicki \& Lightman 1979, Sec. 4.8).\footnote{In normal
astrophysical plasmas where one averages over many electron gyrations
received and emitted power are essentially the same, since the time
scale is set by the duration of one gyration where the electron
approaches and recedes from the observer. Here we only consider the
approaching part of one gyration.}  Half of the emission will be
beamed into a cone of half-opening angle $\phi$ determined
by the beaming cone of synchrotron emission which is
$\phi\sim1/\gamma_{\rm e}\simeq 1^\circ (\gamma_{\rm e}/60)^{-1}$.
This gives a received power of 

\begin{equation} \label{snutheo1}
S_\nu=\frac{1}{2} \gamma_{\rm e}^2 N_{\rm e}^2 P_{\rm e}A^{-1}\nu_{\rm
c}^{-1}\left({\Delta \nu\over\nu_{\rm
c}}\right)\left({\nu\over\nu_{\rm c}}\right)^{1/3}
\end{equation}

where $A=\pi R^2$ is the illuminated area at the ground and $R=\phi
H\simeq 100\,{\rm m}\,(H/60\,{\rm km})(\gamma_{\rm e}/60)^{-1}$. This is the
correct size scale for air showers (see Fig.~\ref{beamshape}).  The
de-coherence of synchrotron radiation due to the shower thickness
limits the validity of the equation to $\Delta\nu\ll\nu_{\rm
max}\sim150$ MHz.
The `dilution factor' $\left({\Delta \nu\over\nu_{\rm
c}}\right)$ in Eq.~\ref{snutheo1} accounts for the fact that
for a bandwidth-limited observation the pulse becomes smeared
out. Incoherence limits the maximum bandwidth to $\Delta\nu\ll\nu_{\rm
max}$. We also take into account that the flux density of synchrotron
radiation actually decreases as $\nu^{1/3}$.

The total density of $e^\pm$ can be roughly estimated as a
function of primary energy $E_{\rm p}$ (see, e.g., Allan 1971):

\begin{equation}\label{neinshower}
N_{\rm e}\simeq {E_{\rm p}\over{\rm GeV}}.
\end{equation}

The integral number of coherently radiating particles around a
characteristic energy is set to be $N_{\rm e}(\gamma_{\rm e})$, with
$N_{\rm e}(\gamma_{\rm e})\sim N_{\rm e}$ around $\gamma_{\rm
e}=\gamma_{\rm e,min}$. At $\nu < 150$ MHz the equivalent flux density
is then predicted to be

\begin{equation}\label{snutheo2}
S_\nu\simeq 32\,{\rm MJy}\;\left({N_{\rm e}(\gamma_{\rm
e})\over N_{\rm e}}\right)^2\left({E_{\rm p}\over{10^{17}\,{\rm
eV}}}\right)^2 \left({\gamma_{\rm e,min}\over 60}\right)^{4/3}
 \left({\Delta\nu\over1{\rm
MHz}}\right) \left({\nu\over50{\rm MHz}}\right)^{1\over3}
\end{equation}
where we kept the shower height fixed at $H=6$ km.

According to Eq.~\ref{fluxconv} this is about 11 $\mu$V m$^{-1}$
MHz$^{-1}$ which is consistent with the empirical formula
(Eq.~\ref{crvoltage}) at 50 MHz and slightly above the values claimed
in the later Haverah Park observations for showers above $10^{17}$ eV.
This shows that, while the derivation presented here is crude, we seem
to be able to account for the level of radio emission observed from
EAS at least within an order of magnitude.

The fall-off of the spectrum beyond 100 MHz (Spencer 1969) could be
explained qualitatively by the loss of coherence at $\nu>\nu_{\rm
max}$ (see Aloisio \& Blasi 2002, for a more involved discussion of
this point).  Once the emitting layer is a multiple of the wavelength,
the waves from coherent regions of size $c/\nu$ will add destructively
with the exception of a small excess layer of order of $c/\nu$. This
effectively reduces the number of contributing particles as $N_{\rm
e,eff}\propto\nu^{-1}$ and we get roughly $S_\nu\propto\nu^{-2}$ times
a smaller correction factor due to the non-flatness of the spectrum.

The claimed E-W polarization is also naturally expected from coherent
geosynchrotron emission since synchrotron emission is intrinsically
highly polarized. For the modest Lorentz factors considered here one
would also expect to see some circular polarization at or below the
percent level. The exact amount will depend on the negative charge
excess and the average electron energy.

Clearly, more sophisticated models have to be developed taking into
account the results of Monte Carlo simulations of showers, the full
electromagnetic wave production, and the shower evolution, curvature,
height, and lateral distribution. However, for the purpose of
understanding the basic EAS radio properties the simple formulation
presented here provides at least an intuitive starting point --
especially for radio astronomers and particle physicists who are used
to think in terms of synchrotron emission.

Of course, one should not discount other emission processes that have
been discussed in the past, such as Cherenkov radiation or
bremsstrahlung. The data are not sufficient to exclude that such
processes could also play a role in certain regimes. For now we can
only state that for primary energies around $10^{17}$ eV and in the
frequency range around 100 MHz, geosynchrotron seems to be sufficient
to explain the observations. Higher statistics, higher time
resolution, more polarization measurements, and multi-frequency data
are urgently needed. It would also be interesting to know whether,
similar to optical fluorescence, there is also a faint isotropic radio
afterglow, e.g. from low-energy electrons, or 'fluorescent'
emission. The effect of the energy (of sometimes macroscopic
dimensions) dumped by one ultra-high energy cosmic ray into the
atmosphere could also lead to some interesting effects, such as radar
(which may actually be FM radio stations) reflections (see Blackett \&
Lovell 1940) or changes in the atmospheric transmission.
 
\section{Detecting EAS radio emission with LOFAR}
\subsection{Basic LOFAR design}
LOFAR, the Low-Frequency Array\footnote{http://www.lofar.org}, is a
new attempt to revitalize astrophysical research at 10-200 MHz with
the means of modern information technology (see, e.g. Bregman
1999). The array is currently in its design phase with first and
significant funding being already available. Construction could start
as early as 2004 with first data being available in the year 2006.
LOFAR is a predecessor to the Square-Kilometer-Array (SKA)
Square-Kilometer-Array (SKA)\footnote{http://www.skatelescope.org} which will
operate in the GHz regime and is foreseen for 2015.

Antenna and receiver technology at these frequencies have become very
simple and cheap which allows one to have a large array and to put
most of the effort into data processing. The basic idea of LOFAR is
therefore to build a large array of about $10^2$ stations of $10^2$
dipoles (at the lower frequencies), all stations distributed in a
spiral configuration with maximum baseline of $\sim$ 400 km. One
quarter of the antennas will be located in a central core of 2 km
diameter. The initial field of view of the dipoles is about $\pi$
steradian, but the telescope will act as a ``phased array'' where the
phasing is done digitally, yielding a maximal resolution of
$1.4^{\prime\prime}$. The received waves are digitized and sent via
glasfiber Internet connections to a central super-cluster of
computers. The total data rates are expected to exceed 10
Tbit/sec. The computer will then correlate the data streams and
digitally form beams (`virtual telescopes') in any desired
direction. The number of beams, eight are currently planned, and the
time to switch from one position to another depends purely on the
computing power. The computer cluster will also take over the
responsibility for modeling ionospheric effects and taking out
interference.

At low frequencies LOFAR has the possibility to permanently monitor a
large fraction of the sky at once. This will be used to look for
astrophysical transients from very short to long timescales -- such as
gamma ray bursts, X-ray binary flares, stellar outbursts, variability
of active galaxies, etc. This will open a completely new window for
radio astronomy. An interesting feature of the LOFAR design is the
possibility to store the entire data stream for a certain period of time (up
to 5 minutes is currently planned). If one detects a radio flare one can
then retrospectively form a beam in the direction of the flare, thus
basically looking back in time and getting very high gain, resolution,
and sensitivity. LOFAR therefore combines the advantages of a low-gain
antenna (large field of view) and of a high-gain antenna (high
sensitivity and background suppression) at low radio frequencies
through its virtual multi-beaming capability. This makes it an ideal
tool to study the radio emission from cosmic ray air showers in an
unprecedented way.

\subsection{Sensitivity and count rates}

The advertised RMS sensitivity of LOFAR is 10 Jy per beam in one
$\mu$sec at 120 MHz and 280 Jy per beam at 30 MHz for 4 MHz
bandwidth. From equations \ref{crvoltage} \& \ref{fluxconv} we know
that at $R=R_0$ and $\nu=120$ MHz the flux density for a $10^{17}$ eV
cosmic ray in 1 $\mu$s is 15 MJy, formally allowing a secure
$1.5\cdot10^6 \sigma$ detection at 120 MHz if the array has enough
dynamic range. For the inner part of the planned array, the so called
`virtual core' of four square kilometers, we know that such an event
would happen roughly once every 12 minutes. Requesting a sure
10$\sigma$ limit, the detection threshold for EAS radio emission could
be reduced to $E_{\rm p,min}\sim2.5\cdot10^{14}$ eV -- provided one
can extrapolate Eq.~\ref{crvoltage} to these energies.  This is
already below the knee and event rates would be up to 90 per second
for LOFAR.

The sensitivities calculated here are of course per beam -- in fact a
beam that is ideally tailored to the geometric wave form of the radio
pulse and a radio-only detection of the pulse would require a lot of
computational effort. This effort would be much reduced if one can
detect pulses already in individual data streams, i.e. from individual
dipoles. The sensitivity calculated above would then be lowered (RMS
increased) by the number of dipoles making up the virtual core, which
is of order 3000. This pushes the minimum primary energy up by roughly
a factor of $\sqrt{3000}\sim50$ to about $E_{\rm p,min}\sim10^{16}$ eV
with events happening roughly every ten seconds in the virtual core.

We can make those estimates a bit more general and accurate, by
starting from a few fundamental assumptions. Let us assume we have $N$
dipoles with a beam of $\Omega_{\rm beam}=4 \pi g^{-1}$ looking at the
sky and a system temperature of $T_{\rm sys}=100$ K. The system
equivalent flux density (SEFD) for one polarization -- the flux
density of a point source producing the same signal in the receiver --
is then


\begin{equation}
{\rm SEFD}_{\rm dipol}={2 k_{\rm B} T_{\rm sys}\over A_{\rm
eff}}={4\pi 2 k_{\rm B} T_{\rm sys}\over g \lambda^2}= 0.1\,{\rm MJy}\;g^{-1}\left({T_{\rm sys}\over100\,{\rm K}}\right)\left({\nu\over55\,{\rm MHz}}\right),
\end{equation}
where we take an effective area of $A_{\rm eff}=\lambda^2/\Omega_{\rm
beam}=g \left({c/\nu}\right)^2/4\pi$ (e.g., Rohlfs \& Wilson
1996). 

This should be compared to the sky background which is dominating at
low frequencies. To estimate the sky background, we have obtained the
Effelsberg 408 MHz survey (Haslam et al. 1982) and convolved it with a
wide beam of $90^\circ$ suitable for single-element antennas. We find
that the average sky temperature in the northern hemisphere is
$\left<T_{\rm sky}\right>=32$ K, with $\left<T_{\rm sky}\right>=37$ K
in the right ascension range 0$^\circ$-180$^\circ$ including the
Galactic plane towards the Galactic Center 
and $\left<T_{\rm sky}\right>=27$ K in the right ascension range
180$^\circ$-360$^\circ$ including the Galactic pole. In the southern hemisphere
one has $\left<T_{\rm sky}\right>=35$ K.

The flux density spectral index ($S_\nu\propto\nu^{\alpha_{\rm r}}$)
at low frequencies is $\alpha_{r}\simeq-0.5\pm0.1$ (e.g., Cane 1979),
which can be verified by comparing the 408 MHz survey map with a 45
MHz survey map (Maeda et al. 1999; P. Reich priv. comm.). Spectral
index variations are rather small and are included in the quoted
error. Thus we have
\begin{equation}
\left<T_{\rm sky}\right>=32(\pm5)\,{\rm K}\;\left({\nu\over408{\rm MHz}}\right)^{-2.5\pm0.1}.
\end{equation}

The corresponding sky SEFD, defined here by exchanging $T_{\rm sys}$
with $T_{\rm sky}$, is then ($T_{\rm sky}\simeq4800$ K at 55 MHz)


\begin{equation}\label{skynoise}
{\rm SEFD}_{\rm sky}\simeq5.6\,{\rm MJy}\;g^{-1}\left({\nu\over{\rm55\, MHz}}\right)^{-0.5}.
\end{equation}

The RMS  noise for an interferometer with efficiency $\eta\sim0.5$ 
is  given by
\begin{equation}\label{RMS}
{\rm RMS}={1\over\eta}{{\rm SEFD}\over\sqrt{N(N-1)\Delta\nu\Delta t}}=11\,{\rm
MJy}\;\left({{\rm SEFD}\over\,5.6\,{\rm
MJy}}\right)\left(\sqrt{N(N-1)}\right)^{-1}
\end{equation}

where we will set $N(N-1)\rightarrow1$ for $N=1$. Note again that for
a bandwidth-limited (unresolved) pulse we have $\Delta t=1/\Delta\nu$
and the noise is independent of the bandwidth.  To get the
signal-to-noise ratio for an array of $N$ dipoles we simply divide the
expected cosmic ray radio flux density from Eqs.~\ref{fluxconv} \&
\ref{crvoltage} by the RMS,

\begin{eqnarray}
{\rm SNR}&=&7.7\cdot g\,\left({\Delta\nu\over{\rm 16\, MHz}}\right) \left({E_{\rm
p}\over 10^{17} ~{\rm eV}}\right)^2
\left({\sin\alpha\,\cos\theta\over \sin45^\circ\,\cos30^\circ}\right)
\,\nonumber\\
&\times&\exp \left({-R\over {\rm 110\,m}}\right)^2 \left({\nu\over55 ~{\rm
MHz}}\right)^{-1.5}\left({{\rm SEFD}\over\,5.6\,{\rm
MJy}}\right)^{-1}\left(\sqrt{N(N-1)}\right)
\end{eqnarray}
This is valid in the 100 MHz regime. The SNR increases {\em linearly}
with bandwidth until the pulse is resolved. We assume that the
air shower is spatially unresolved by a single dipole, otherwise the SNR will 
not increase with the gain of the dipole antennas.

\begin{figure}
\centerline{\psfig{figure=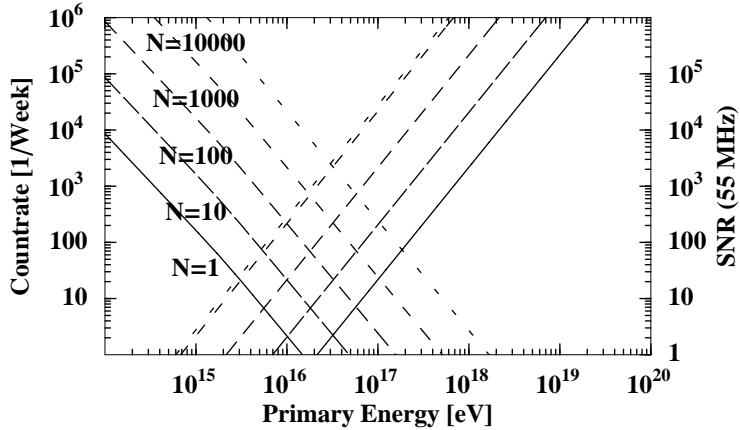,width=0.7\textwidth,bbllx=3.2cm,bblly=22.2cm,bburx=11.3cm,bbury=26.7cm}}
\caption[]{\label{snr}Signal-to-noise (right axis, lines slanted to the 
right) for radio detections with $N$ dipoles and expected count rates
(left axis, lines slanted to the left) of cosmic rays as a function of
cosmic ray energy. The intersection of the two sets of lines with the
x-axis delimits the theoretically useful cosmic ray energy range for an
array of these dimensions. It is assumed that the antennas have a gain
of $g=3$ and are densely packed with an assumed cosmic ray collecting
area of only $(c/\nu)^2$ for each antenna.  The system temperature is
$T_{\rm sys}=300$ K and the (dominating) sky background is calculated
using Eq.~(\ref{skynoise}). The bandwidth is 16 MHz for
bandwidth-limited pulses at 55 MHz. The SNR does not increase beyond
$N=1000$ because the dipoles fall outside the beamed emission of the
air shower. The calculation assumes a zenith angle of $30^\circ$, a
geomagnetic angle of 45$^\circ$, and an offset of 110 m from the
shower core. For the count rates we have formally considered only air
showers where the shower core intersects the effective area of one
dipole ($\sim30$ m$^2$). Since the air shower is larger than this
effective area, configurations with $N<100$ would actually see much
higher count rates.}
\end{figure}

Figure \ref{snr} shows the expected SNR and count rates for various
antenna array configurations as a function of primary energy. One can
see that the minimum detectable cosmic ray energy for a single dipole
is a few times $10^{16}$ eV, similar to what was estimated above.

On the other hand one can ask what the maximum detectable primary
particle energy is. This is mainly limited by the maximum event rate,
since radio sensitivity is not a major issue here. If we
conservatively assume that the detectable radio beam on the ground at
high energies is about 1 km$^2$ and we have about $10^2$ stations in
an array like LOFAR outside the virtual core, we have an effective
area of order 100 km$^2$. Considering one event per year a still
reasonable detection rate, we could see cosmic rays up to $10^{20}$ eV
with LOFAR, i.e. up to the GZK cut-off. This of course assumes that
the emission is beamed. Should one find a detectable isotropic
component (e.g., radar reflection or radio ``fluorescence'') in the
radio band, one could greatly improve this and possibly utilize the
entire size of the array with roughly $10^5$ km$^2$. This would bring
one up to the $4\cdot10^{21}$ eV cosmic rays -- way beyond the GZK
cut-off and an order of magnitude above what has been observed so
far. While this possibility is highly speculative at the moment, we
note that for such energetic events single dipoles instead of entire
stations would be more than enough. Hence, a reconfigured LOFAR design
could easily be applied to a dedicated particle array (e.g. Auger) and
indeed approach these energies.

Figure \ref{snr} shows useful limits in terms of count rates for
densely packed dipole configurations together with the expected
sensitivities. An important feature here is that the count rates are
computed for the primary beam set by a single dipole, while the
sensitivities are calculated for a `virtual beam' formed out of all
dipoles. A single LOFAR-like station with about 100 dipoles would be
useful already for the energy range $10^{15}-10^{17}$ eV which makes
this technology also interesting for current air shower experiments in
this range, such as KASCADE (Klages et al. 1997), provided self-made
interference can be dealt with. Such an experiment, nicknamed
LOPES\footnote{http://www.mpifr-bonn.mpg.de/staff/hfalcke/LOPES}
(LOFAR Prototype Station), is fully funded and currently underway. A
first set of antennas is expected to be operational in 2003 at the
KASCADE site with useful data expected in 2004. At higher energies
densely packed radio arrays are not necessary and one can cover a
large area with a few antennas only.

\subsection{Detection strategies for LOFAR}
The current design of LOFAR calls for the inclusion of a transient
monitor. This will be a piece of software that, in connection with the
online buffering, detects astrophysical transient events. It is clear
that a program to detect radio emission from extensive air showers
(hereafter REAS) will benefit from, help, and interfere with this
transient monitor. In any case, the basic requirements for the
hardware and the software protocols to detect transient phenomena are
already available so that the usage of LOFAR as an astroparticle array
does not require any major redesign. From the considerations in the
previous section it is also clear that in order to build an effective
monitor for astrophysical transients one needs to understand (and
eliminate) REAS.

In principle REAS should produce a number of clearly distinguishable
features: 

\begin{itemize}
\item bursts are short, the pulse duration could be around
10 ns but faint afterglows cannot be excluded
\item the pulse is broad-band
\item the emission is produced in the near-field and the wavefront is
curved
\item the emission is highly linearly polarized in E-W direction and
weakly circularly polarized with a fixed sign
\item bursts are localized to a few stations only
\end{itemize}

What does this practically mean? Suppose we digitize the incoming
waves with a rate of 65 MHz, corresponding to a sampling time of 15
ns. The pulse would be smeared over 250 ns due to bandwidth smearing
in a 4 MHz window, corresponding to 17 bins.  Hence continuously
comparing a running average with a 20 bin window from individual
dipoles with their mean should quickly allow detection of radio pulses
from $E_{\rm p}>10^{16-17}$ eV. This could be done as part of the
transient monitoring. Upon detection of such a pulse at multiple
dipoles of a station within a coincidence window of about 10 $\mu$s
the data stream around this interval would be dumped and fed into a
post-processing algorithm -- this could happen in principle several
times a minute. Alternatively, one could also consider using an
external trigger from particle detectors.

Given enough signal-to-noise for energetic cosmic rays the arrival
times could possibly be determined from the un-averaged data to almost
the sampling time of 15 ns. From the pulse amplitudes as a function of
antenna location one can determine the shower core. The arrival times
determine the wave front curvature and inclination, allowing one to
determine the shower direction. The light travel time across the
virtual core (2 km) is about 6.7 $\mu$s and with an accuracy of 15 ns
one could in principle locate arrival directions to within
$0.4^\circ$.

In general the wavefront curvature is not only due to the emission
being produced in the near-field, for coherent emission it also
contains information about the shape of the shower itself. Both
effects should have a relatively well determined functional form. With
a good guess of what this curvature should be, one could predict
arrival times at more distant antennas (from the shower core) to
detect even fainter signals.  To first order one could approximate the
emission as being coherent on cylinders intersecting an inclined plane
and one would sum the signals from individual dipoles in an ellipse on
the ground after applying the appropriate time shift. In principle one
should thus be able to devise a self-calibration-like scheme where a
shower model is iteratively adjusted until it produces maximum
correlation at all antennas for the detected pulse. This would be
equivalent to forming an `adaptive beam' in the shower direction,
where the beam would depend not only on the position on the sky but
also on the shower geometry and height. Such a software could perhaps
be generalized to locate the position of arbitrary nearby bright radio
bursts, e.g. to localize sources of man-made interference. With the
gained sensitivity of such an iteratively formed adaptive beam one could
then try to determine further pulse properties, such as polarization,
spectrum, and shower shape.

Especially the pulse shape should be of major interest, since so far
the REAS pulse shape has not been convincingly resolved. In the
current design the maximum bandwidth of LOFAR is 32 MHz which is split
into eight 4 MHz bands. Some proposals have been made to increase this
bandwidth even further to 64 MHz or more. In any case, interference
will prevent one from using the full bandwidth. Still, one could try
to sample the full bandwidth at various frequencies and reconstruct
the pulse shape in the Fourier domain from only a few frequency
windows with a ``CLEAN'' algorithm (H\"ogbohm 1974). This would be
similar to reconstructing images from snapshot data of an array with
sparsely filled aperture as is commonly done in radio astronomy. The
achievable time-resolution with LOFAR could then be between 15 and 32
ns depending on the actually allowed maximum bandwidth.

An additional more involved program could be to reconstruct the cosmic 
ray track for bright events and then retrospectively form a beam from
the {\it entire} array focusing at the shower maximum to look for
faint, isotropic afterglow emission.

The computational load for all these programs should be manageable
since one only needs to work on 10 $\mu$s worth of data for the
central core, which corresponds to roughly 1 kB per antenna, i.e. 10
MB for the entire data set of $\sim10^{4}$ low-frequency
antennas. The initial radio-only detection which is based on running
averages would be part of the transient monitor or general data
quality control routines. The actual data analysis program could be
partially run as a filler program: low-energy cosmic rays which are
frequent would be analyzed only if time is available otherwise they
would be ignored. On the other hand, obviously bright pulses would
have to be processed with very high priority. This way, one would
never have to waste computational time with LOFAR, since it can
always be run as an air shower detector, but one can also
ignore a lot of the faint cosmic rays if the array is used for other
purposes.

\subsection{Cooperation with particle detector arrays} 
Air showers are commonly observed by directly detecting the fast
leptons hitting particle counters on the ground. An ideal situation
would be to combine such a particle array with the radio capabilities
of LOFAR. For example, the particle detectors can be directly used to
trigger LOFAR. Especially for low-energy cosmic rays around and below
the knee, such particle detectors could provide a valuable first
guess for the REAS self-calibration routine to detect the radio emission in
the first place. A blind-search for cosmic rays that are too weak to
be detected by individual antennas in the data stream would place a
major computational burden on the LOFAR computer and at the lowest
energies would probably be hopeless.

Moreover, the particle detectors will be crucial in the early phase to
verify the correlation between certain types of radio bursts and
EAS. In addition, a combined particle and LOFAR array would allow
cross-calibration. Since there is relatively little experience still
on how to relate radio pulse properties to cosmic ray energies and
composition, one needs to first `train' the LOFAR algorithm with the
established results of particle detectors. In the final phase the
combination of LOFAR and particle detectors should allow one to obtain
a significantly improved calibration for the combined array with
respect to the stand-alone arrays, because the radio and particle
detectors measure the shower at two very different stages in its
evolution.

A few groups are currently developing a concept to build a large
particle array,
named``SKYVIEW''\footnote{http://skyview.uni-wuppertal.de/}, in the
western part of Germany and perhaps parts of the Netherlands. The idea
is to combine particle detectors in groups of three or four and place
them on public buildings or schools (see, e.g., Meyer 2001). Each
group would look for local coincidences from EAS and report every
detection, tagged with precise GPS times, via Internet to a central
processing station. Since public buildings and schools are quite
frequent in the heavily populated area of western Germany (Ruhrgebiet)
-- roughly every kilometer -- a patchy but giant air shower array
could be built up rather easily. In addition, the schools could
actively use the local air shower stations for their own experiments,
thus providing a great public outreach and science education
opportunity. Each station would mainly consist of a few flat boxes
with scintillator material and photo multipliers, a computer, and a
few cables. If appropriately shielded, a few of these particle array
stations could also be installed near LOFAR stations: each particle
array station is easily transportable and relatively cheap ($<5000$
EUR). While first funding for prototypes of this project have been
approved, the time line of SKYVIEW is unclear and depends strongly on
future funding.

In addition, once we understand REAS better, one could consider
upgrading such a particle array with simple dipoles, receivers, A/D
converter units, and small data buffers. Upon detection of an
energetic event by the particle detectors the radio data could be sent
via Internet to a data processor (e.g., the LOFAR computing center)
and used to also detect and evaluate the radio emission.

Alternatively, as mentioned above, the LOFAR concept can be applied to
already existing arrays such as KASCADE with a single prototype
station, as in the LOPES experiment. Since one is only interested in
short-term bursts and triggering is done by the already
well-calibrated particle array, the computational and data transfer
load can be reduced to a bare minimum. One needs about 100 dipoles
with fast A/D converters, online storage, and a fast Ethernet
connection. Each dipole would produce about 2 kB of data per burst for
100 MHz sampling. Fourier transforming, filtering and correlation of
the total dataset of 200 kB can be done rather quickly on a powerful
workstation. This experiment will be crucial to properly calibrate any
LOFAR air shower data.

Finally, if the technique is well-established, one may think of
equipping larger cosmic ray arrays, e.g., Auger which is located in
a radio-quiet zone, with radio antennas. Here one antenna per station
would be sufficient and data could be transmitted using mobile-phone
technology.

\subsection{Requirements for LOFAR}
What are the requirements for the LOFAR design following from these
considerations? A lot can be done already with the current design and
a few additional things could be put on the wish-list: 

\begin{itemize}
\item use of maximum bandwidth to increase time-resolution: at least 32
MHz -- better is 64 MHz --  or at least simultaneous observations at widely separated 
frequencies
\item high dynamic range for each antenna, i.e. 14 bit
sampling or more
\item simultaneous usage of the low- and and high-frequency part of
the array
to get the spectrum and an improved time-resolution
\item buffer for un-averaged data with the possibility to transmit the
transient buffer data also from stations outside the virtual core
\item incorporation of a CR detection algorithm into the transient monitor
and inclusion of flexible scheduling with varying priorities
(depending on CR energy) for the data analysis
program
\item allow for possible upgrade of LOFAR station with particle detectors (power
outlet, Internet connection, space for four 1m$^2$ boxes ``inside the
fence'')
\end{itemize}

As Green et al. (2002) have shown a significant bit-depth (more than
8-bit) is really a crucial requirement. 

\subsection{Scientific gain from LOFAR}
Finally, after having outlined what the prospects for cosmic ray air
shower detections with LOFAR are, we briefly want to summarize what
the scientific perspective of such an undertaking is.  The first
objective will be to study REAS themselves and understand the basic
process leading to the radio emission in the first place. LOFAR offers
several orders of magnitude higher sensitivity and count rates in
comparison to earlier experiments. So far a major uncertainty has been
the high beaming, leading to largely varying radio pulse as a function
of distance from the shower core. For the first time we will now get
fully spatially resolved maps of individual radio bursts. Since the
radio emission is produced by the fast electrons moving through a very
homogeneous magnetic field, the radio emission should accurately
reflect the shower development, especially the electron distribution
in the shower maximum, if measured properly.

LOFAR will thus allow one to relate the measured radio properties of
EAS to energies and composition of the primary cosmic ray
particles. Additional information from the radio spectrum and
time-resolved pulses could be obtained at higher frequencies, but this
may have to wait for the construction of telescopes like the SKA.

In a second step LOFAR will then be able to very accurately measure
the cosmic ray spectrum from $2\cdot10^{14}$ eV to $10^{20}$ eV.  An
interesting aspect will be the composition of CRs around the knee and
up to $10^{18}$ eV and the possible clustering of ultra-high-energy
cosmic rays. Here LOFAR could easily compete with all current
arrays. The wide energy range of LOFAR is a unique feature coming from
it being a scaled array with many different baselines. Typical
particle arrays usually have a single baseline length (or grid
constant) thus narrowing the observable energy range. Therefore LOFAR
would also be sensitive to unexpected changes in air shower properties
that have possibly been missed so far, e.g. multiple or very patchy
air showers.  The long baselines of LOFAR might help, for example, to
detect the Gerasimova-Zatsepin effect (Gerasimova \& Zatsepin 1960;
Medina-Tanco \& Watson 1999), which predicts widely separated showers from
photo-disintegration of comic ray nuclei near the sun. This effect allows
one to determine cosmic ray masses.

Another, very interesting aspect will be the correlation of the cosmic
ray flux around $10^{18}$ eV with the low-frequency radio map of the
Galaxy that LOFAR is going to produce with unprecedented
clarity. Because of diffusion in the Galactic magnetic field, charged
cosmic rays should usually appear homogeneous on the sky with some
possible asymmetries due to magnetic field gradients that can be
derived from radio maps. This is, however, not true for neutrons which
would travel on straight lines and could make up a few percent of the
incoming cosmic rays. For $10^{18}$ eV neutrons the Lorentz factor is
about $\gamma\sim10^9$ and the lifetime of neutrons becomes of order
$10^{12}$ sec. This allows a free path length before decay of order 10
kpc, corresponding to the distance to the Galactic Center.  Small,
localized excesses in the cosmic ray flux would thus help to pinpoint
individual sources of high-energy neutrons (see for example Tr\"umper
1971). Such an excess has already been claimed towards the Galactic
Center and the Cygnus region (Hayashida et al. 1999). In a similar
vein LOFAR, with its ability to detect variable transient sources,
like stellar coronae and winds, neutron stars and supernovae, would
also be able to correlate outbursts from such sources to possible
``neutron bursts'', i.e. temporary and spatially constrained excesses
of the cosmic ray intensity. In this sense LOFAR could actually do
real neutron astronomy (see, e.g., Biermann et al. 2001).

\section{Detecting Neutrinos with LOFAR}

Although as noted previously, high energy neutrino fluxes are 
quite uncertain at present, there is considerable interest in
developing techniques for large area detectors which may constrain
or directly measure neutrino interactions in the PeV to EeV
energy regime and beyond. There are several different scenarios in 
which LOFAR may have a unique corner of sensitivity to neutrino
interactions. The fundamental requirement is that there be some
intervening radio-transparent matter to produce a neutrino interaction
and the resulting cascade. Such material can be found
in the earth just below the array, in the
atmosphere above it, or even in the lunar regolith when the moon 
is in view of the array. In this section
we will describe these different possibilities in general terms.

\subsection{Interactions from below}

At energies of about 1 PeV, the earth becomes opaque to neutrinos at
the nadir. For higher energies, the angular region of opacity
grows from around the nadir till at EeV energies, neutrinos can
only arrive from within a few degrees below the horizon. The interaction
length at these energies is of order 1000 km in water, so such
neutrinos have a significant probability of interacting along a
$\sim 100$ km chord. If the interaction takes place within several
meters below the surface in dry, sandy soil, the resulting cascade will
produce coherent Cherenkov radiation up to microwave frequencies
which can refract through the surface and may be detected as a 
surface wave, depending on the antenna response. The flux density
expected for such events (cf. Saltzberg et al. 2001) is
\begin{equation}
S_{\nu} \simeq 12 {\rm~MJy}  
\left ( {R \over {\rm 1~km}} \right )^{-2}
\left ( {E_{c} \over 10^{18}{~\rm eV}} \right )
\left ( {\nu \over 200 {~\rm MHz}} \right )^2
\end{equation}
where $E_c$ is the cascade energy and $R$ the distance to
the cascade.\footnote{Note that in this case the neutrino energy is not
necessarily equal to the cascade energy $E_c$, because for
the typical deep-inelastic scattering interactions that occur
for EeV  neutrinos, only about 20\% of the energy is put into
the cascade, while the balance is carried off by a lepton.
For electron neutrinos, the electron will rapidly interact
and add its energy to the shower, but for muon or tau neutrinos,
this lepton will generally escape undetected (although the
tau lepton will itself decay within a few tens of km at 1 EeV).}
The Cherenkov process
weights these events strongly toward the higher frequencies,
though events that originate deeper in the ground will
have their spectrum flattened by the typical $\nu^{-1}$ behavior of
the loss tangent of the material.

A similar process leads to
coherent transition radiation (TR; cf. Takahashi et al. 1994) 
from the charge
excess of the shower, if the cascade breaks through the
local surface. TR has spectral properties that make it more
favorable for an array at lower frequencies such as LOFAR:
it produces equal power per unit bandwidth across the
coherence region.
The resulting flux density for a neutrino cascade breaking the
surface within the LOFAR array, observed at an angle of 
within $\sim 10^{\circ}$
from the cascade axis, is (cf. Gorham et al. 2000):
\begin{equation}
S_{\nu,TR} (\theta \leq 10^{\circ}) ~\simeq~ 2~{\rm MJy} 
\left ( {R \over {\rm 1~km}} \right )^{-2}
\left ( {E_c \over 10^{18} ~{\rm eV}} \right )^{2}~.
\end{equation}
The implication here is that, if LOFAR can retain some
response from the antennas to near-horizon fluxes, the
payoff may be a significant sensitivity to neutrino events
in an energy regime of great interest around 1 EeV,
or even significantly below this energy depending on the
method of triggering.

\subsection{Neutrino interactions in the atmosphere}

Neutrinos can themselves also produce air showers. The primary
difference between these and cosmic-ray-induced air showers
is that their origin, or first-interaction point, can be 
anywhere in the air column, with an equal probability of
interaction at any column depth. Neutrino air showers can
even be locally up-going at modest angles, subject to the earth-shadowing
effects mentioned above.

Detection of such events is identical to detection of cosmic-ray-induced
air showers, except for the fact that sensitivity to events
from near the horizon is desirable, since these will be most easily
distinguished from cosmic-ray-induced events. Beyond a zenith
angle of $\sim 70^{\circ}$ cosmic-ray radio events will be 
more rare, and those that are detected in radio will be distant. 
The column depth of
the atmosphere rises by a factor of 30 from zenith to horizon; thus
cosmic ray induced air showers have their maxima many kilometers away at
high zenith angles. Neutrino showers in contrast may appear close
by, even at large zenith angles.

Of particular interest is the possibility of observing ``double--bang''
(Learned \& Pakvasa 1995) 
tau neutrino events. In these events, a $\nu_\tau$ interacts first,
producing a near-horizontal air shower from a deep-inelastic 
hadronic scattering interaction. The tau lepton escapes with
of order 80\% of the neutrino energy, and then propagates an
average distance of $50 E_{\tau}/(10^{18}~{\rm eV})$~km 
before decaying and producing (in most cases) another shower
of comparable energy to the first. Detection of both cascades
within the array boundaries of LOFAR would provide a unique
signature of such events. And in light of the SuperKamiokande
results indicating $\nu_\mu \rightarrow \nu_{\tau}$ oscillations,
it is likely that neutrinos from astrophysically distant sources
would be maximally mixed, leading to a significant rate of
$\nu_{\tau}$ events.

Another interesting possibility is to look for the radio emission from
up-going air showers that is reflected off the lower ionosphere at
low frequencies, within the 10-30 MHz band during daytime observations.
Since this band is in any case dead to astronomical sources during the day,
one could attempt to optimize sensitivity for impulsive events
during this fraction of the time. One needs to first study the
coherence that might be retained on reflection at these
frequencies and also whether the signature of such a reflection
could be uniquely identified.

\subsection{Lunar regolith interactions}

There is an analogous process to the earth-surface layer cascades
mentioned above which can take place in the lunar
surface material (the regolith). In this case the cascade takes
place as the neutrino nears its exit point on the moon after having
traversed a chord through the lunar limb. This process, first
suggested by Dagkesamansky \& Zheleznykh (1989) 
is the basis of several searches
for diffuse neutrino fluxes at energies of $\sim 10^{20}$ eV
(Hankins et al. 1996; Gorham et al. 1999, 2001) using large
radio telescopes at microwave frequencies. Based on the 
simulations for these experiments (Alvarez M\~uniz \& Zas 1997,1998; 
Zas, Halzen \& Stanev 1992)
and confirmation through
several accelerator measurements (Gorham et al. 2000; Saltzberg et al. 2001), 
the expected flux density from
such an event at about 1 attenuation-length depth in the
regolith can be roughly estimated as
\begin{equation}
S_\nu~=~ 50\,{\rm Jy}\; \left ( {E_{c} \over 10^{20}{~\rm eV}} \right )
\left ( {\nu \over 200 {~\rm MHz}} \right )^2~.
\end{equation}
Note here that the flux density is far lower than for air shower events,
but the two should not be compared, since the lunar regolith events are
coherent over the entire LOFAR array, and originate from a small, known
angular region of the sky (the surface of the moon). Thus their detectability
depends on the sensitivity of the synthesized beam, depending on the
ability of the system to trigger on bandwidth-limited pulses.

Transition radiation events may also be detectable in
a similar manner, as noted above. For TR from events that break the
lunar surface, the resulting pulse differs from a Cherenkov
pulse because it is flat-spectrum. Because TR is strongly
forward beamed compared to the Cherenkov radiation from the moon,
we estimate that the maximum
flux density for this case, at an angle of $\sim 1.5^{\circ}$
from the cascade axis, 
is about a factor of 20 higher than at $\sim 10^{\circ}$. At earth
the implied flux density for LOFAR is:
\begin{equation}
S_{max,TR} (\theta \simeq 1.5^{\circ}) ~\simeq~ 40\,{\rm Jy}\;
\left ( {E_c \over 10^{20} ~{\rm eV}} \right )^{2}~.
\end{equation}
Although this channel does not provide a higher flux density than
the Cherenkov process, it is a flat spectrum process that may provide
more integrated flux across the LOFAR band.

These pulses are essentially
completely bandwidth-limited prior to their entry into the ionosphere,
with intrinsic width of order 0.2 ns.
Dispersion delay in the ionosphere will of course significantly
impact the shape of any pulse of lunar origin. This will
limit the coherence bandwidth for a VHF system. The 
dominant quadratic part of the dispersion gives an overall
delay 
\begin{equation}
\tau_{ion} ~=~ 1.34 \times 10^{-7} {N_{\rm e} \over \nu^2}
\end{equation}
where $\tau_{ion}$ is the delay in seconds at frequency $\nu$ (in Hz)
for ionospheric column density $N_{\rm e}$ in electrons per m$^{2}$. For typical 
nighttime values of $N_{\rm e} \sim 10^{17}$ m$^{-2}$
the zenith delay at 200 MHz is 330 ns, and the differential 
dispersion is of order 3 ns per MHz, increasing at lower frequencies as
$\nu^{-3}$. For bandwidths up to even several
tens of MHz for zenith observations, and perhaps a few MHz
at low elevations, the pulses should remain bandwidth-limited. However,
coherent de-dispersion will be necessary to accurately
reconstruct the broad-band pulse structure.

Although the problem of coherent de-dispersion is a difficult one, a
LOFAR system may have an edge in sensitivity over systems operating at
higher frequencies, under conditions where the intrinsic neutrino
spectra are very hard. This is due to the fact that the loss tangent
of the lunar surface material is relatively constant with frequency
(Olhoeft and Strangway 1976), and thus the attenuation length
increases inversely with frequency.  This means that a lower frequency
array may probe a much larger effective volume of mass than the higher
frequencies can. At 200 MHz, the RF attenuation length should be of
order 50 m or more, compared to 5-7 m at 2 GHz.  When this larger
effective volume is coupled with the larger acceptance solid angle
afforded by the broader RF beam of the low-frequency Cherenkov
emission, the net improvement in neutrino aperture could well
compensate the loss of sensitivity at lower energies by a large
margin.

It is also worth noting here that these lunar regolith observations
are distinct from other methods in high energy particle detection,
in that they do require the array to track an astronomical target,
and can and will make use of the synthetic beam of the entire array.
This is because, although the sub-array elements should be used
for the detection since they will have a beam that covers the entire
moon, the Cherenkov beam pattern from an event of lunar origin
covers an area of several thousand km wide at earth, and is 
thus broad enough to trigger the entire array. Post-analysis
of such events can then localize them to a few km on the lunar surface,
and provide opportunities for more detailed reconstruction of
the event geometry.

\section{Conclusions}
While the investigation of the radio emission from extensive air
showers has lain dormant for a rather long time there is enough
information available that suggests that this field could be
revived. The properties of these radio pulses from cosmic ray air
showers are all consistent with it being coherent geosynchrotron
emission from electrons and positrons in the air shower. This process
is basically unavoidable and hence the radio emission should directly
reflect the shower evolution of the leptonic component of cosmic ray
air showers if properly measured. 

Because the emission is highly beamed, a key to successful usage of
the radio emission is the rather new possibility to build digital
telescopes that combine a large field of view with the ability to
form virtual beams retrospectively in the direction of transient
events. For this reason, the planned radio array LOFAR (and possibly
also the SKA) will become a very efficient cosmic ray detector which
is sensitive to high-energy cosmic rays at all energies from
$\sim10^{14}$ to $10^{20}$ eV. The calibration and accuracy could be
further improved by combining this digital radio technology with
existing or upcoming air shower arrays consisting of particle counters
on the ground (KASCADE/LOPES,Auger).

The combination of radio techniques and particle counters should
provide a unique tool to study the energy spectrum and composition of
cosmic rays over a broad range rather efficiently, simultaneously
probing a parameter space never combined in a single array. Moreover,
at energies around $10^{18}$ eV neutron astronomy would, for the
first time, become possible. A large radio array like LOFAR could also
be used to search for radio emission from neutrino induced showers in
the air or from the lunar regolith, possibly opening a new window to
the universe. Hence,  digital radio telescopes could provide a
significant technological advantage for astronomy and astroparticle
physics. 

\medskip
{\bf Acknowledgment:} Many people have contributed with their
suggestions and comments to this paper. We particularly want to thank
Tim Huege, Alan Roy, Peter Biermann, Karl Mannheim, Ger de Bruyn,  and
the participants of the LOFAR workshop in Dwingeloo, May 2001, for
intensive discussions on this and related topics. We thank Patricia
Reich for providing us with a properly smoothed version of the 408 MHz
sky survey.  We are also grateful to the editor, Alan Watson, an
anonymous referee, and Harold Allan for helpful comments on the
manuscript. Ralph Spencer provided some useful insight into the
details of the early radio experiments. We thank Walter
Fu\ss{}h\"oller for helping to prepare some of the figures. This work
has been performed in part at the Jet Propulsion Laboratory,
California Institute of Technology, under contract with the
U.S. National Aeronautics and Space Administration.  Funding was also
provided in part by the German Ministry of Education and Research
(BMBF), grant 05 CS1ERA/1 (LOPES).

\end{document}